\documentclass[aps,prl.reprint,twocolumn]{revtex4} 
\usepackage{epsfig}
\usepackage{graphicx}
\usepackage{amsmath}
\usepackage{amsfonts}
\usepackage{amssymb}
\usepackage{ulem}
\usepackage{cancel}
\usepackage{graphicx}
 \usepackage{wrapfig}
\usepackage[cp1251]{inputenc}
\usepackage{epsfig}
\usepackage{graphicx}
\usepackage{amsmath}
\usepackage{amsfonts}
\usepackage{amssymb}
\usepackage{graphicx}
 \usepackage{wrapfig}
\usepackage{textcomp}
\usepackage{pgf,pgfarrows,pgfnodes,pgfautomata,pgfheaps,pgfshade}
\usepackage{amsmath,amssymb}
\usepackage[colorlinks=true, linkcolor=blue]{hyperref}
\newcommand{{\sign}}{\rm sign}

\begin{document}
\title{
Thermoelectric transport in junctions of Majorana and Dirac channels
}
\author{Dmitriy S. Shapiro$^{1,2,3,4}$} \email{shapiro.dima@gmail.com} \author{D. E. Feldman$^{5}$}
\author{Alexander D. Mirlin$^{6,7,8}$} \author{Alexander Shnirman$^{7}$}
\affiliation{$^1$V. A. Kotel'nikov Institute of Radio Engineering and Electronics, Russian Academy of Sciences, Moscow 125009, Russia}
\affiliation{$^2$L. D. Landau Institute for Theoretical Physics, Russian Academy of Sciences, Moscow 117940, Russia}
\affiliation{$^3$Dukhov Research Institute of Automatics (VNIIA),  Moscow 127055, Russia}
\affiliation{$^4$Moscow Institute of Physics and Technology,  Dolgoprudny 141700,  Russia}
\affiliation{$^5$Department of Physics, Brown University, Providence, Rhode Island 02912, USA}
\affiliation{$^6$Institut f\"ur Nanotechnologie, Karlsruhe Institute of Technology, 76021 Karlsruhe, Germany}
\affiliation{$^7$Institut f\"ur Theorie der Kondensierten Materie, Karlsruhe Institute of Technology, 76128 Karlsruhe, Germany}
\affiliation{$^8$Petersburg Nuclear Physics Institute,  St.Petersburg 188300, Russia}

\begin{abstract}
We investigate the thermoelectric current and heat conductance in a chiral Josephson contact on a surface of a 3D topological insulator, covered with superconducting and magnetic insulator films. The contact consists of two junctions of Majorana and Dirac channels next to two superconductors. Geometric asymmetry results in a supercurrent without a phase bias.   The interference of Dirac fermions causes oscillations of  the electric and heat currents with    an unconventional  period $2\Phi_0=h/e$ as functions of the Aharonov-Bohm flux.   Due to the gapless character of Majorana modes, there is no threshold for the thermoelectric effect and the current-flux relationship is non-sinusoidal. Depending on the magnetic flux, the direction of the electric current can be both from the hot to cold lead and vice versa. 
\end{abstract}

\maketitle

\section{Introduction} 
A Majorana fermion is simply the real or imaginary part of a complex fermion. At first sight this implies that no meaningful distinction exists between systems of complex and Majorana fermions. However, it is more practical and much more conventional to use the language of complex fermions for normal metals and many other systems. On the other hand, Majorana fermions provide a natural description for various topological materials. The simplest example is the Kitaev chain \cite{Kitaev}. Its low-energy degrees of freedom are two Majorana excitations at the chain's ends. Two-dimensional topological materials bring richer examples of Majorana physics. For example, Majorana edge modes \cite{rmp-nonab} are expected in several candidates states \cite{review-5-2,php} for the quantum Hall effect at the filling factor $5/2$.

If a Majorana system is in contact with a system of complex fermions then a natural question concerns transformations between the two types of fermions when the systems exchange electrons. The simplest version of that question involves electron tunneling \cite{tn1,tn2,tn3}. A more interesting setting is a Y-junction of Majorana modes that merge into a Dirac quantum channel. Such junctions can be build on a surface of a 3D topological insulator (TI) ~\cite{FuKane3DTI, FuKaneMachZehnder}.

It has long been known that 1D charge-neutral  Majorana fermions can exist as subgap  edge modes of  2D chiral $p$-wave topological superconductors \cite{qi2011, Alicea2012}.  An $s$-wave superconductor (SC) can also give rise to such modes  in a partially gapped  hybrid structure with a superconducting film on a surface of a TI. A splitted film that hosts an SC-insulator-SC interface on top of a 3D TI supports a gapped non-chiral 1D  Majorana mode  while   an SC/ferromagnet  junction supports a  gapless chiral  one ($\chi$MM) ~\cite{FuKane3DTI, FuKaneMachZehnder}.    Recently topological superconductivity and Majorana 1D edge modes were reported in an anomalous quantum Hall insulator/SC heterostructure  \cite{He} and in a single atomic Pb layer on a magnetic Co/Si(111) island \cite{Menard}.

A magnetic domain wall on top of a TI hosts a chiral Dirac mode ($\chi$DM). Combinations of such domain walls with SC/magnet junctions allow the implementation of novel quantum devices.
The simplest example is a Y-junction of Majorana and Dirac modes. Other proposals include
the Mach-Zehnder \cite{FuKaneMachZehnder, AkhmerovMachZehnder}, Fabry-P\' erot \cite{LawFabryPerot, ButtikerFabryPerot}, and Hanbury-Brown Twiss \cite{HBT} quantum interferometers.  
In our work \cite{ShapiroShnirmanMirlin} we introduced a 3D TI-based chiral Josephson contact.

The previous work has focused on electric transport in the above devices. In the present paper we extend this line of research to thermoelectric and thermal transport. 
Our motivation comes from the question about the nonequilibrium state that forms, if two Majorana modes with different temperatures fuse into a Dirac mode.
We focus on the setup from Ref. \onlinecite{ShapiroShnirmanMirlin} and derive analytical expressions for the thermoelectric and heat currents in the presence of the magnetic field through the normal region.    Note that thermal transport between a lead and a 1D Majorana mode has been studied in Ref. \onlinecite{thermo-DF}. The case of  localized Majorana bound states has been studied in Ref. \onlinecite{Ramos}.

\begin{figure}[h] 
 	\includegraphics[width=\linewidth]{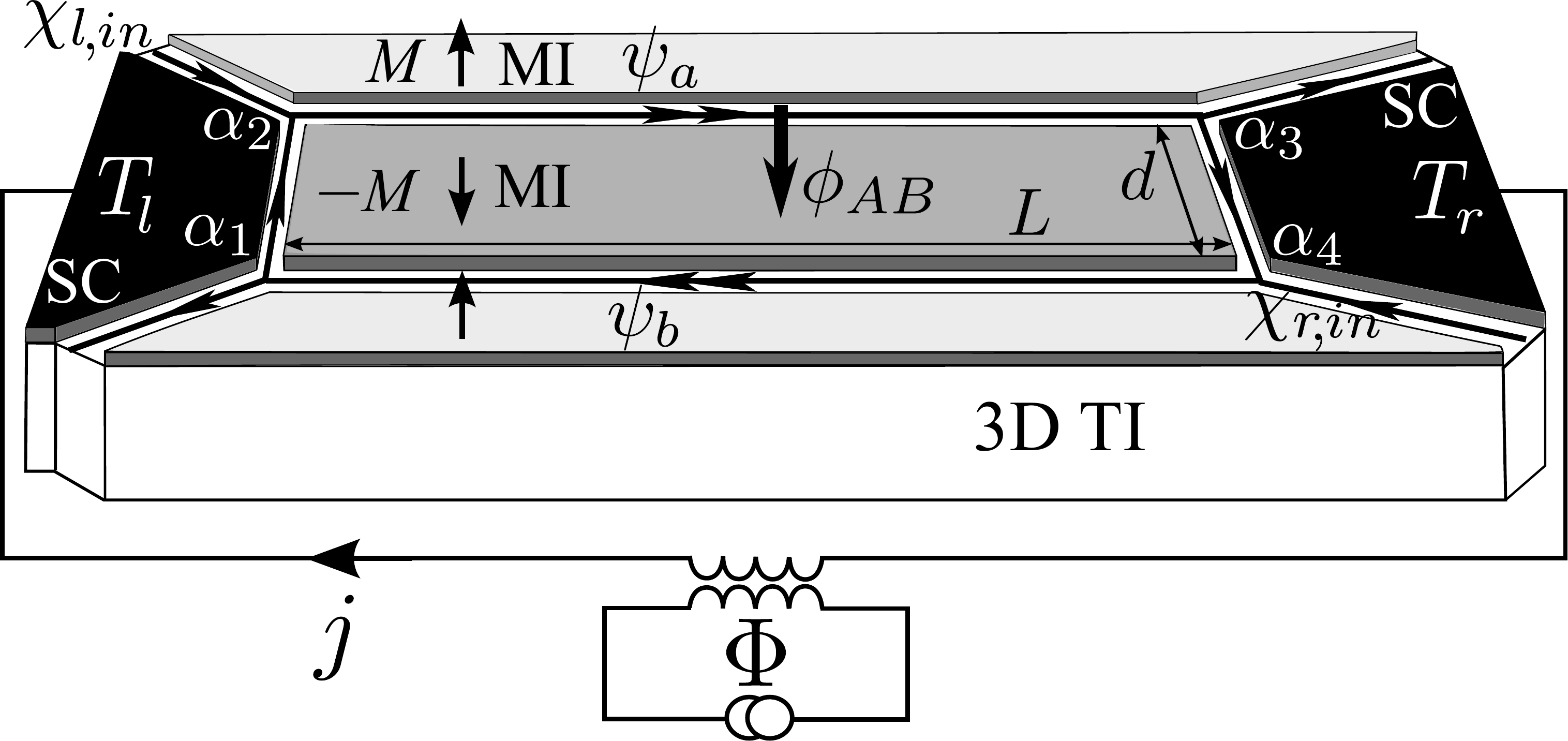} 
 	\caption{ A chiral Josephson junction on a surface of a 3D topological insulator  (3D TI). The   lines   with a single arrow surrounding black  superconducting films (SC)  stand for gapless Majorana fermion  channels  $\chi_{r,l}$ and the arrows show chiralities. Superconducting electrodes have different temperatures $T_l$ and $T_r$. The light and dark gray areas are magnetic insulators (MI) which induce exchange fields of the opposite polarizations and energy gaps $\pm M$. Magnetic domain walls support chiral charged modes  $\psi_{a,b}$ marked  by double arrows.   A magnetic flux $f$ in the $-M$ region induces the  Aharonov-Bohm phase  $\phi_{AB}=\pi f/\Phi_0$.} \label{setup}
 \end{figure}

Our device is shown in Fig. \ref{setup}. It can be understood as a Fabry-P\' erot interferometer made of four chiral Y-junctions. Each junction converts neutral Majorana fermions into charged Dirac particles. The charge is supplied by a superconductor.
 The device is a relative of  a   quantum-Hall-based Josephson junction  with a gapped superconductor and 
 a quantum Hall bar in the normal region \cite{Ostaay,Zyuzin}. In such a structure, the supercurrent is carried   by chiral edge states. A recent experimental realization of a quantum Hall junction involved  molybdenum-rhenium contacts mediated by a $\mu$m sized graphene bar encapsulated
 in boron nitride \cite{Amet}. An important feature, common to our setting and the quantum Hall device, is the spatial separation of electrons and holes in Andreev pairs due to the spatial separation of the chiral transport channels. 
One consequence of such splitting is a `single-electron' Aharonov-Bohm periodicity in the transport behavior: All transport quantities are periodic in the magnetic flux through the gray region of Fig. \ref{setup} with the period $2\Phi_0=h/e$. 
For comparison,  S/N/S junctions, based on quantum spin-Hall (or 2D TI) films \cite{Tkachov, LMAY, Baxevanis} or two-channel nanowires \cite{Mironov}, exhibit even-odd transitions between the $\Phi_0$-  and $2\Phi_0$-periodicities.  
  The heat transport and interference effects in thermally biased 2D TI-based Josephson junctions   have been studied in Refs. \onlinecite{thermo1-DS,thermo2-DS}.

Below we compute the thermal and thermoelectric currents. The time reversal symmetry is broken by the  magnetic film. As a consequence, the inevitable geometric asymmetry of the junction results in a nonzero electric current even in the absence of a temperature gradient, a phase difference between superconductors, and an Aharonov-Bohm flux.  
 The thermoelectric effect requires particle-hole asymmetry. This asymmetry is due to the Aharonov-Bohm effect.
Our results reveal a significant difference of thermal transport in the setup of Fig. \ref{setup} from the threshold-like transport in a conventional S/N/S junction with gapped leads. 
The thermoelectric current oscillates as a function of the Aharonov-Bohm flux, and, consequently, as a function of  the interferometer area. The oscillation amplitude is geometry dependent. We find the maximal current of the order of  $eE_{Th}/\hbar\sim e/\tau$, where $E_{Th}$ is the Thouless energy and $\tau$ is the electron travel time through the device,   if  the temperatures of the leads satisfy   $T_{l}\gg E_{Th} \gg T_{r}$ or $T_{r}\gg E_{Th} \gg T_{l}$. The $2\Phi_0$-periodic heat conductance oscillates from zero to  one half of the heat conductance quantum. The maximum heat conductance agrees with what is expected for a fully transparent junction of chiral Majorana channels \cite{Gnezdilov}. Note that the experimental measurement of quantized thermal conductance has recently been accomplished in the integer \cite {iqhe-thermo} and fractional \cite{fqhe-thermo} quantum Hall effect.

The width $d$ and length $L$ of the normal region, bounded by two counter-propagating charged $\chi$DMs, are much longer than the coherence lengths $\xi$ of the induced superconductivity. Hence, the Thouless energy, proportional to the inverse travel time through the interferometer, is much lower  than the superconducting proximity gap and the magnetic exchange gap $E_{Th}\ll \Delta,M$. 
 We assume that the temperatures of the incoming Majorana modes are below those gaps. 
We will mostly focus on the case of a much higher exchange than superconducting gap $M\gg\Delta$. 
 In this case all contributions to the Josephson current arise from the 1D Dirac channels and do not involve the 2D band between the superconducting leads. Indeed, we expect no contributions to the Josephson effect from the energies $E>M\gg\Delta$. The gray 2D area exhibits insulating behavior for the energies below $M$ and the tunneling through the insulator is suppressed due to its large size $L\gg\xi$.
 Another assumption is that the superconducting leads are   large and have a constant  chemical potential which crosses the Dirac point. This means that the DC Josephson effect in this contact is $2\pi$-periodic because the fermion parity is not conserved. The unconventional non-equilibrium $4\pi$-periodic component, predicted in Refs. \cite{Kitaev,Jiang11,FuKane,1,2,Ioselevich} for localized zero-energy Majorana bound states \cite{Mourik},  is suppressed in our device.

Since we only consider 1D physics, we ignore phonons in the bulk. Phonons are not expected to have much effect on the electric current. They do contribute to the thermal conductance. We are only interested in the oscillating contribution from topological modes. One can isolate it experimentally in a setting,  where two hot Majorana modes are brought to a cold device.

\begin{figure}[h]  
 \includegraphics[width=\linewidth]{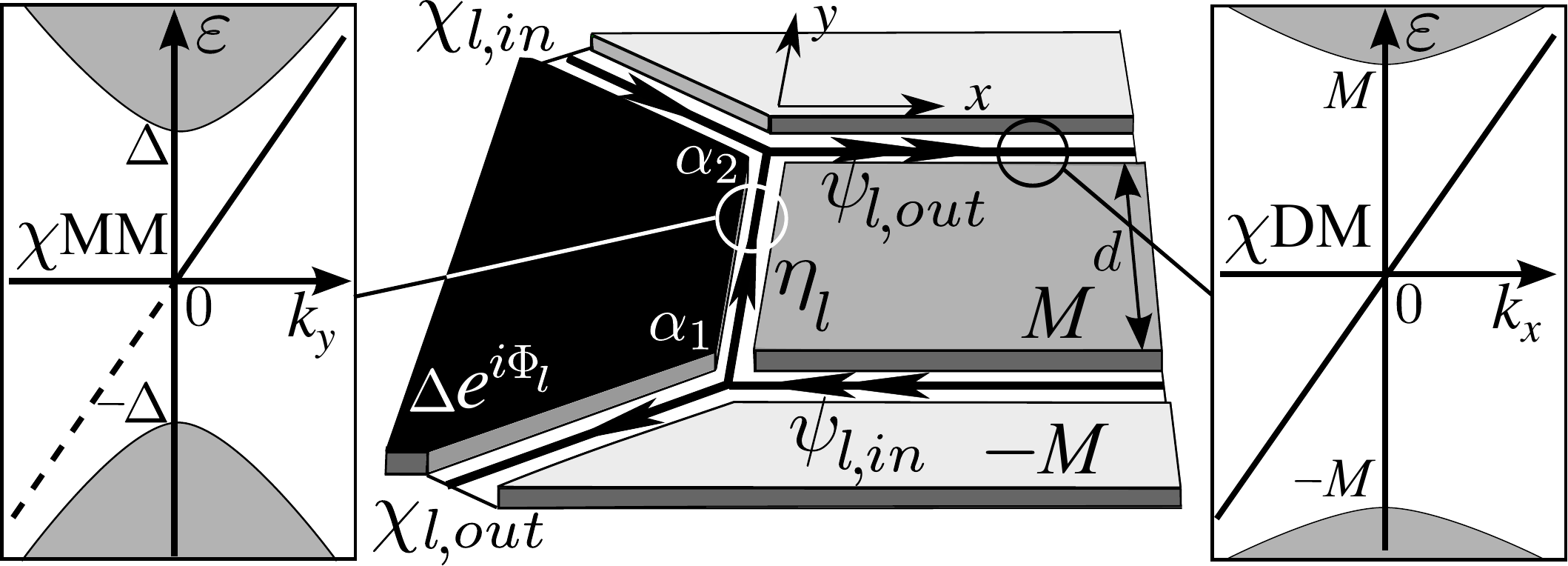}
 	\caption{The structure of the chiral Dirac-Majorana 1D contact formed by two Y-junctions. Film  of a superconductor with the phase of the order parameter $\Phi_{l}$ and magnetic insulators of the opposite magnetizations  induce proximity $\Delta$ and exchange gaps $\pm M$ on the 2D helical surface. The  boundaries between the superconductors and magnets support the chiral charged modes $\psi_{in,out}$ and the neutral modes $\chi_{in,out}$ and $\eta_l$.} \label{contact}
 \end{figure}

\section{  Dirac and Majorana 1D liquids}
	The mean-field Hamiltonian of the 2D structure introduced in Fig. \ref{setup} reads
    \begin{multline}
    H=\frac{1}{2}\int dx dy \Psi^+ h \Psi\ , \,  h=iv\tau_z \mathbf{z} \cdot (\boldsymbol \sigma \times \boldsymbol{\nabla})+
    \\+\tau_0 \sigma_z M(x,y)+(\tau_+ \Delta(x,y) +\tau_-\Delta^*(x,y))\sigma_0, \label{H}
    \end{multline}
    where $\boldsymbol{\sigma}$ and $\boldsymbol{\tau}$ are the Pauli matrices in the spin and Nambu spaces. The spinor $\Psi = [\psi_\uparrow,\psi_\downarrow,\psi_\downarrow^+, -\psi_\uparrow^+]^T$ contains field  operators of free electrons and holes  on the  surface of the topological insulator.
     The helical states of the 2D surface   are described by the Rashba Hamiltonian with the Fermi velocity $v$ and the chemical potential $\mu=0$ crossing the Dirac point. The superconducting $s$-wave pairing potential is given by $\tau_+\sigma_0\Delta(x,y)$ while the exchange field of the magnetic insulator films is described by $\tau_0\sigma_zM(x,y)$ term. The black areas of right (left) SC contacts have    $\Delta(x,y)=\Delta e^{-i\Phi_{r,l}}$. In the normal region filled with  magnetic films the magnetization is perpendicular to the 2D surface and changes its sign: in the light gray  regions the induced exchange gap $M(x,y)=M$ and in the dark  gray rectangle $M(x,y)=-M$. Both $M$ and $\Delta$ are real.
     
     An effective 1D Hamiltonian  of  a Majorana   mode, like the one marked by single arrow   in Fig.  \ref{contact},  was derived by Fu and Kane \cite{FuKaneMachZehnder}. This derivation is based on a solution of  the 2D Bogolyubov-de Gennes 
equation. Below we review that solution for the mode $\eta$ which  connects Y-junctions 1 and 2 in Fig. \ref{contact} and propagates along the SC/magnet interface at $x=0$. In the SC region (the $x<0$ half plane), there is an  $s$-wave SC pairing potential given by $\Delta(x,y)=\Delta e^{i\Phi_{l}}\theta(-x)$, 
   while at $x>0$ the magnetic film induces the  exchange gap $M(x,y)=-M\,\theta(x)$.
There exists a 1D solution  of the Bogolyubov-de Gennes equation $h\xi_{k_y}=\varepsilon_{k_y}\xi_{k_y}$ such that the wave function decays exponentially in the directions, normal to the SC/magnet interface as $\sim\exp[-|(\theta(x)M-\theta(-x)\Delta)x|/(\hbar v)]$ and is a plane wave with the momentum $k_y$ along the boundary. This 1D chiral mode with the dispersion relation 
  \begin{equation}
   \varepsilon_{k_y}= - {\sign }(M) v {k_y}
  \end{equation} 
is {\it nondegenerate} within the gap, i.e., for $\varepsilon_{k_y} < {\rm min}(\Delta,M)$, but continues to exist 
also for higher energies.
The eigenvectors are self conjugate, $\xi_{k_y}=\sigma_y\tau_y\xi_{-{k_y}}^*$, which is consistent with the 
   the fact that the field $\Psi$ is self charge conjugate, $\Psi=\sigma_y\tau_y \Psi^*$.
   Hence,  the Bogolyubov quasiparticle operator  
    \begin{equation}
   \label{eq:chip}
   \eta_{k_y}=\int dxdy (\xi_{k_y} (x,y))^\dag\cdot \Psi(x,y)
    \end{equation}
   is real, $\eta_{k_y}=\eta^+_{-k_y},$ and   describes  a chiral Majorana mode. 
   
The normal region with Dirac modes is confined by domain walls where the    magnetization sign changes (the horizontal   lines marked by double arrows in  Fig. \ref{contact}). 
To derive the effective 1D Hamiltonian for those modes from the 2D Hamiltonian in the Nambu space (\ref{H}), we set $\Delta(x,y)=0$ and focus on the mass term  $M(x,y)=M\,{\sign}(y)$ at $y\approx 0$.

The  eigenvalues $\varepsilon_{k_x}$  of  the Bogolyubov-de Gennes Hamiltonian are now {\it doubly} degenerate in contrast to the  case of $\chi$MM.  We denote two  orthogonal degenerate eigenstates as $\zeta_{e,k_x}$ and   $\zeta_{h,k_x}$ and  associate them with electrons and holes. Their wave functions are related via the charge conjugation constraint as  $\zeta_{h,{k_x}}=\sigma_y\tau_y  \zeta_{e,-{k_x}}^{*}$. 
The dispersion relation is the same as for the  Majorana channels: $\varepsilon_{k_x}= - {\sign }(M) v {k_x}$. The difference from the  neutral mode consists in the existence of two independent excitations of the same energy $\varepsilon_{k_x}$ in the Nambu space: an electron of momentum ${k_x}$  and a hole of momentum $-{k_x}$. In terms of  Bogolyubov operators this is the  Dirac gapless mode   described by a complex field. 
	In the second quantization language, the electron and hole operators are given by
   	  \begin{equation}\label{eq:psiep}
   	\psi_{e,k_x}=\int dxdy (\zeta_{e,k_x} (x,y))^\dag\cdot \Psi(x,y)\ ,
   	\end{equation}
   	 and 
   	\begin{equation} \label{eq:psihp}
   	\psi_{h,k_x}=\int dxdy (\zeta_{h,k_x} (x,y))^\dag\cdot \Psi(x,y)\ .
	\end{equation}
   They are not independent since $\psi_{h,k_x}=\psi_{e,-k_x}^+$ due to the charge conjugation constraints for $\Psi(x,y)$ and $\zeta_{e,{k_x}},\zeta_{h,{k_x}}$. In what follows we do not use $\psi_{h,e}$ and instead introduce the field $\psi_{k_x}$ such that $\psi_{e,k_x}=\psi_{k_x}$ and $\psi_{h,k_x}=\psi_{-k_x}^+$. 
   	
   	At this point we are in the position to write down effective 1D Hamiltonians for free Majorana and Dirac particles.   The secondary quantized $\Psi$-operators of these  1D modes  are: 
\begin{equation}\Psi_M(x,y)=\int\frac{dk_y}{2\pi}\xi_{k_y}(x,y)\eta_{k_y}
\end{equation}
 for $\chi$MM and
\begin{equation}    \Psi_D(x,y)=\int\frac{dk_x}{2\pi} \left( \zeta_{e,k_x} (x,y)\psi_{k_x}+\zeta_{h,k_x}(x,y)\psi_{-k_x}^+  \right ) 
\end{equation}
 for $\chi$DM.
   Integrating out the $y$ coordinate in the Bogolyubov-de Gennes Hamiltonian yields the effective Hamiltonians of the Majorana modes 
   \begin{equation} H_M={\sign}(M)\frac{iv}{2}\int\!  \eta(y)\partial_y\eta(y) dy
\end{equation}
   and the Dirac modes
  \begin{equation} 
  H_D={\sign}(M)iv\int\!  \psi^+(x)\partial_x\psi(x) dx,\end{equation}
 where we introduced the 1D operators   \begin{equation} \eta(y)=\eta^+(y)=\int\frac{d{k_y}}{2\pi}\eta_{k_y} e^{i{k_y}y}\end{equation}
 and \begin{equation}\psi(x)=\int\frac{d{k_x}}{2\pi}\psi_{k_x} e^{i{k_x}x},\end{equation} 
 which describe   coherent propagation of neutral and charged fermions through 1D guiding channels with the Fermi velocity $v$.   For $M>0$ the chiralities of the 1D modes are shown by the arrows in Figs. \ref{setup} and \ref{contact}.  The factor $1/2$ in the Majorana Hamiltonian $H_M$ reflects the fact that the negative and positive energy excitations in the $\chi$MM are not independent. In other words, the lower   branch of the dispersion $\varepsilon_{k_y}=-  {\sign }(M) v {k_y} $  at ${k_y}<0$ is redundant (it is shown as a dashed line in the left inset of Fig. \ref{contact}). The coefficient $1/2$ implies that a neutral Majorana fermion   carries only a half of the heat current of a Dirac mode at the same temperature  so that the ballistic heat conductance $G_0$ of a single $\chi$MM  is one half of the heat conductance quantum: \begin{equation}
 	G_0=\frac{1}{2}\frac{\pi^2 k_{\rm B}^2 T}{3h},
 \end{equation}
  where $T$ is the temperature.

\section{Scattering in a Majorana-Dirac   contact }
 The normal region includes a rectangular magnetic film (dark gray area in Fig. \ref{setup}) of the length $L$ and the width $d$. $L$ and $d$ exceed significantly both the  SC and magnetic  coherence lengths $d,L \gg \hbar v/\Delta, \hbar v/M$.  Four Y-junctions are in the corners of the film.   A single Y-junction is  formed by two Majorana and one Dirac channels (Fig. \ref{contact}). The angles between the channels as well as other microscopic details are not necessarily the same in different Y-junctions. 
 
We start with the calculation of the scattering matrix 
describing two nonidentical Y-junctions  shown in Fig. \ref{contact} (see~\cite{ShapiroShnirmanMirlin}). This scattering matrix describes the coupling between $\chi$MMs on the SC/magnet interfaces and two 1D Dirac modes. Specifically, it provides a relation between the operators of incoming and outgoing electrons and holes $\psi_{in,out},\psi_{in,out}^+$ of $\chi$DM (horizontal lines marked by double arrow)  and $\chi_{in,out}$ of semi-infinite neutral $\chi$MMs (lines marked by single arrow). 
   The 1D modes, described by the wave functions $\xi$ and $\zeta_{e,h}$ are spin-nondegenerate and have in-plane spin textures.  Hence, the conversion between Majorana and Dirac modes  in Y-junctions is accompanied by spin rotation. Thus, scattering in a Y-junction involves a geometric Berry phase, which is encoded in the phase $\alpha$ below. The calculation of $\alpha$ for a given geometry is straightforward.

   Scattering in the  upper and lower Y-junctions in Fig. \ref{contact} is described by the $S_{out}$ and  $S_{in}$ matrices   which were found  in Refs. \onlinecite{AkhmerovMachZehnder, FuKaneMachZehnder} 
   \begin{equation}\label{eq:Sin}
   \begin{bmatrix}\eta_{l,out} \\ \\
   \chi_{l,out}
   \end{bmatrix}=
   S_{in,\alpha_1}
   \begin{bmatrix}\psi_{l,in} \\ \\ \psi_{l,in}^+
   \end{bmatrix}, \, \, \, \begin{bmatrix}\psi_{l,out} \\ \\ \psi_{l,out}^+
   \end{bmatrix}= S_{out,\alpha_2}
   \begin{bmatrix}\eta_{l,in} \\ \\
   \chi_{l,in}
   \end{bmatrix}
   \end{equation}
  Note that the operators in (\ref{eq:Sin}) correspond to the incoming and outgoing scattering states rather than    to free plane waves of (\ref{eq:chip}), (\ref{eq:psiep}), and (\ref{eq:psihp})  \cite{Blanter}.

Let us assume first that $\Phi_{l}=0 $ in the electrode. A non-zero $\Phi_{l}$ will be included in a final expression for the $S$-matrix  by means of  a gauge transformation of Dirac $\psi$-operators.  The matrix $S_{in,\alpha_1}$ involves the phase difference $\alpha_1$ between  an electron and hole converting into two Majorana fermions. The matrix $S_{out,\alpha_2}$ involves a phase $\alpha_2$ accumulated under merging  two  Majoranas into a Dirac fermion. The structure of  $S_{out} $ is related to that of $S_{in}$ by a time-reversal transformation \cite{AkhmerovMachZehnder}: $S_{out}=S_{in}^T\label{s-out}.$  The expression for the
   $S_{in}$-matrix of the lower Y-junction is
   \begin{equation}
   \quad S_{in,\alpha_1}=\begin{bmatrix}
   1/\sqrt{2} &&  1/\sqrt{2}\\ \\ i/\sqrt{2}  && -i/\sqrt{2}
   \end{bmatrix}\begin{bmatrix}
   e^{i\alpha_1} &&  0\\ \\ 0  && e^{-i\alpha_1}
   \end{bmatrix}.\label{s-in}
   \end{equation} 
Note that whereas the phase $\alpha_1$ can be easily gauged out if the Y-junction is considered on its own, it becomes important once several Y-junctions are combined into a circuit.

In  Ref. \cite{ShapiroShnirmanMirlin} the symmetry of four Y-junctions ($\alpha_i=\alpha$) was assumed. In this paper we consider an arbitrary set of the phases $\alpha_i$.  We will see that this modifies the current-phase relation  in such a way that a nonzero current may flow at a zero external phase bias $\Phi$, like in Josephson $\varphi$-junction devices  \cite{Buzdin, Goldobin, Sickinger}.

   We proceed by matching the Majorana operators $\eta_{l,in}$ and $\eta_{l,out}$ at a given energy $\varepsilon$ as $\eta_{l,in, \varepsilon}=e^{i k_\varepsilon}\eta_{l,out,\varepsilon}$. The dynamic phase $k_\varepsilon=\varepsilon d/v$ is accumulated by a Majorana excitation during the propagation from the lower to  upper Y-junctions, separated by the distance $d$. The full $S_{\alpha_1,\alpha_2}$-matrix of the   contact, acting on $(\psi_{in, \varepsilon}, \ \chi_{in, \varepsilon}, \ \psi^+_{in, -\varepsilon})^T$, can be found after the  exclusion of $\eta$ from Eqs.~(\ref{eq:Sin})  and is defined by the equation
   \begin{multline}
   \begin{bmatrix}
   \psi_{l,out, \varepsilon} \\ \\
   \chi_{l,out, \varepsilon} \\ \\
   \psi^+_{l,out, -\varepsilon} \\
   \end{bmatrix}
   = \label{s-matr-setup} \\
   =\begin{bmatrix}
   \frac{1}{2} e^{i k_\varepsilon+i( \text{$\alpha_1 $}+\text{$\alpha_2 $})} & \frac{i e^{i \text{$\alpha_2 $}}}{\sqrt{2}} & \frac{1}{2} e^{i k_\varepsilon-i( \text{$\alpha_1 $}-\text{$\alpha_2 $})} \\ \\
   \frac{i e^{i \text{$\alpha_1 $}}}{\sqrt{2}} & 0 & -\frac{i e^{-i\text{$\alpha_1 $} }}{\sqrt{2}} \\ \\
   \frac{1}{2} e^{i k_\varepsilon + i( \text{$\alpha_1 $}-\text{$\alpha_2 $})} & -\frac{i e^{-i \text{$\alpha_2 $} }}{\sqrt{2}} & \frac{1}{2} e^{i k_\varepsilon - i( \text{$\alpha_1 $}+\text{$\alpha_2 $})} \\
   \end{bmatrix}
   \begin{bmatrix}
   \psi_{l,in, \varepsilon} \\ \\
   \chi_{l,in, \varepsilon} \\ \\
   \psi^+_{l,in, -\varepsilon} \\
   \end{bmatrix}.
   \end{multline}
 To account for a non-zero SC phase $\Phi_{SC}$ of an electrode  (colored black in Figure \ref{contact}), we employ the transformation $\psi\to e^{i\Phi_{SC}/2}\psi$. For the left contact in Fig. \ref{setup} this yields 
\begin{equation}
   S_l=C (-\Phi_l)S(\alpha_1,\alpha_2)C(\Phi_l),
\end{equation}
   while for the scattering matrix for the right contact it gives
   \begin{equation}
   S_r=C^{-1}(\Phi_r)S(\alpha_3,\alpha_4)C(\Phi_r).
   \end{equation}
Here we have introduced an auxiliary matrix
  \begin{equation}C(\Phi_{SC})=\begin{bmatrix}
     e^{i\Phi_{SC}} & 0 & 0 \\ \\
  0 & 1 & 0 \\ \\
   0 & 0 & e^{-i\Phi_{SC}} \\
   \end{bmatrix} .
 \end{equation}
The above $S_{l,r}$-matrices describe    partial Andreev reflection in spinless 1D Dirac channels and the creation of excitations in neutral Majorana modes. The Andreev part of this process is accompanied by a Cooper pair absorption in an SC electrode.

   	We transform the $S_{l,r}$-matrices acting on $\psi_{l,in}, \psi_{l,out }$ ($\psi_{r,in}, \psi_{r,out}$) on the left (right) ends of the Dirac channels into new matrices $\tilde S_{l,r}$, acting on the operators $\psi_{a},\psi_{b}$ in the geometric centers of the 1D channels. The $S$ and $\tilde S$ operators are related by a phase shift by the sum of the dynamical phase $\frac{\varepsilon L}{2\hbar v}$, accumulated by an electron of energy $\varepsilon$ over the  distance $L/2$,  and  an Aharonov-Bohm phase. For the upper $a$-arm  the relation of the scattering matrices can be deduced from the equation
\begin{equation}   	
   \psi_{a,\varepsilon}=e^{i\frac{\varepsilon L}{2\hbar v}+i\phi_{AB}/4}\psi_{l,out,\varepsilon}.
\end{equation}
 We assume here that the same Aharonov-Bohm phases are accumulated on each portion of the Dirac channels of the same length. 
  The scattering matrices, acting on the $\psi$-operators in the centers of the channels, take the form
   	\begin{equation}
   	\tilde S_{l,r}=C\left(\frac{\phi_{AB}}{4}\right)D\left(\frac{\varepsilon L}{2 v} \right) S_{l,r} D\left(\frac{\varepsilon L}{2 v} \right) C\left(\frac{\phi_{AB}}{4} \right), \label{sLR}
   	\end{equation}
   	where $\phi_{AB}$ is the total Aharonov-Bohm phase.
   	The dynamical  phases are encoded in (\ref{sLR}) via the matrix 
\begin{equation} 
D\left(\frac{\varepsilon L}{2 \hbar v} \right)=\begin{bmatrix}
   	e^{i\frac{\varepsilon L}{2\hbar v} } & 0 & 0 \\ \\
   	0 & 1 & 0 \\ \\
   	0 & 0 & e^{i\frac{\varepsilon L}{2\hbar v} } \\
   	\end{bmatrix} .
\end{equation}   	
   	  The difference of the above expression from the $C$-matrix is that the first and third diagonal components   coincide: the dynamical phases are equal for an electron of the energy $\varepsilon$ and a hole of the energy $-\varepsilon$.
   	We can exclude $\alpha_i$ from the diagonal terms of the $S$-matrices by redefining the superconducting phase bias $\Phi$  and the Aharonov-Bohm phase $\phi_{AB}$. To do that we introduce the phases  
\begin{equation}
	\phi_l= \frac{\alpha_1+\alpha_2}{2} , \quad \phi_r= \frac{ \alpha_3+\alpha_4}{2}
 \end{equation} 
   	and 
\begin{equation}
   		\varphi_0=\frac{\Phi_l+\Phi_r}{2}.
\end{equation}   		 
We can always shift both superconducting phases by the same constant. It will be convenient to shift them so that 	
\begin{equation}\varphi_0=\frac{\alpha_2-\alpha_1+\alpha_4-\alpha_3}{4}.
\end{equation}
 With this choice we find 
\begin{equation}   S_{l}=C\left(\phi_{l}-(\Phi + \varphi)/2 \right) S_0 C\left(\phi_{l}+(\Phi + \varphi)/2 \right) 
\end{equation}
and 
\begin{equation}
  	S_{r}=C\left(\phi_{r}+(\Phi + \varphi)/2 \right) S_0 C\left(\phi_{r}-(\Phi + \varphi)/2 \right),
\end{equation}  	
   	where $S_0\equiv S_{\alpha_{1}=\alpha_{2}=0}$  (\ref{s-matr-setup}), $$\Phi=\Phi_l-\Phi_r$$ is the SC phase bias, and the phase shift  
\begin{equation}  
\varphi= \frac{1}{2}(\alpha_1-\alpha_2-\alpha_3+\alpha_4). \label{varphi}
\end{equation} 
   	It follows from this representation of $S_{l,r}$ and $\tilde S_{l,r}$ that the superconducting phase cannot be gauged out by the Aharonov-Bohm phase. We also observe that $\phi_{l,r}$ and $\phi_{AB}$ enter the $C$-matrices in the same way, and hence, $\phi_{l,r}$ can be gauged out by redefining the  total Aharonov-Bohm phase  as 
   	\begin{equation}
   	\phi_{AB}\to \phi_{AB}+2(\phi_l+\phi_r)= \sum\limits_i^4\alpha_i+\phi_{AB}.
   	\end{equation}

\section{Josephson current}
 In our formalism the operators of Dirac fermions are expressed as  linear combinations of uncorrelated field operators $\chi_{l}\equiv \chi_{l,in}$ and $ \chi_{r}\equiv \chi_{r,in}$ of incident Majorana modes. The latter are characterized 
 by the Fermi distribution functions 
 \begin{equation}
 n_{l,r}(\varepsilon)=\frac{1}{2}\left(1-\tanh\frac{\varepsilon}{2T_{l,r}}\right), \label{nLR}
 \end{equation} 
 i.e., 
 \begin{equation}\langle\chi_{\varepsilon,i}^\dag\chi_{\varepsilon,j}^{\phantom\dag}\rangle = v^{-1} \delta_{i,j} n_i(\varepsilon),\label{chiLR}
  \end{equation} 
 where the Fourier transformed operators 
 \begin{equation} \chi_{\varepsilon,i}=\chi_{-\varepsilon,i}^\dag=\int \chi_i(t) e^{i\varepsilon t}dt,  
 \end{equation} 
and the index $i=l,r$ stands for the left and right incident modes and  $v^{-1}$ is the density of states in the $\chi$MM channels.  We assume  $k_{\rm B}=1$ everywhere  and recover it in the final expressions.
 The linear spectrum of 1D Dirac modes  means that the chiral current is proportional to the charge density $j_{a,b}=-ev\rho_{a,b}$  and, hence, the   current  $j$ is given by the integral over energies 
 \begin{equation}	
 j= \int \frac{d\varepsilon}{2\pi \hbar} (- e  v) (\langle\psi_{a,\varepsilon}^+\psi_{a,\varepsilon}\rangle-\langle\psi_{b,\varepsilon}^+\psi_{b,\varepsilon}\rangle)\ .  
  \label{j-1}
 \end{equation}
Here $\psi_{a,\varepsilon}$ and $\psi_{b,\varepsilon}$ are the electron operators in the centers of the Dirac channels and  the positive direction of the current is defined from the left to the right.

The $S$-matrices  are  used to express the Dirac fermion operators in Eq. (\ref{j-1}) in terms of the incoming Majorana modes 
\begin{multline}
 \psi_{a,\varepsilon}= \frac{i \sqrt{2}e^{\frac{1}{4} i (\frac{2 L \varepsilon}{v} +\phi- \Phi-\varphi )}  }{ 1+2e^{i   \phi }\cos(\Phi+\varphi)+e^{2i   \phi}  -4 e^{i  (\phi-\varphi_\varepsilon) }}\times\\ \left[  \left(1+e^{i (\Phi +\varphi  +\phi )}-2 e^{i (\phi-\varphi_\varepsilon)}\right) \chi _l- \right. \\
\left.  -2ie^{  i\phi-i\frac{1}{2}\varphi_\varepsilon}\sin\left(\frac{\Phi+\varphi+\phi}{2}\right) \chi _r \right]   \label{psi-a-0}
\end{multline}
 with \begin{equation}\phi=\phi_{AB}+\sum\limits_i\alpha_i. \label{phi}\end{equation}
   Due to the symmetry between the $a$ and $b$ arms we get a similar expression for $\psi_{b}$ with $\Phi\to -\Phi, \varphi\to-\varphi$ and the interchanged  $\chi_{l}$ and $\chi_r$. This expression for Dirac operators is a straightforward generalization of that from Ref. \onlinecite{ShapiroShnirmanMirlin}  to  our asymmetric setup:  (i) the sum of the phases $\alpha_i$ shifts the total  Aharonov-Bohm phase (\ref{phi})  
   and (ii)   the phase $\varphi$,  introduced in (\ref{varphi}), shifts the external superconducting phase bias $\Phi$.

With the use of   the expressions for Dirac operators in N-region  (\ref{psi-a-0}) we   obtain  that the  current   can be represented as 
 \begin{equation}
j=j_t+j_\Phi \label{jFull}
  \end{equation} 
where $j_t$ is induced by the temperature gradient and $j_\Phi$ generalizes the Josephson current from Ref. \onlinecite{ShapiroShnirmanMirlin} to a two-temperature situation. The two contributions read as
 \begin{equation}
 j_t=\int \frac{d\varepsilon}{2\pi \hbar} (- e) \frac{n_l(\varepsilon) - n_r(\varepsilon)}{2}J_{t,\varepsilon}, \label{j-t}
  \end{equation}
  and
  \begin{equation}
  j_\Phi=\int \frac{d\varepsilon}{2\pi \hbar} (- e) \frac{n_l(\varepsilon) + n_r(\varepsilon)}{2}J_{\varepsilon}.  \label{jPhi}
  \end{equation} 
Thermoelectric part (\ref{j-t}) is given by a rapidly convergent   integral due to the factor $(n_l-n_r)$.
If  $T_l=T_r$ then $j_t=0$ and the total current (\ref{jFull}) is given by the  Josephson term $j_\Phi$. We calculate $j_\Phi$  in this section and analyze $j_t$ in the next one.

 The spectral current $J_{\varepsilon}$, entering into  $j_\Phi$, reads 
  \begin{equation}
  J_\varepsilon=\frac{\sin \varphi_\varepsilon \sin(\Phi+\varphi) }{1+\left(\frac{\cos  \phi +\cos (\Phi+\varphi)}{2}\right)^2-(\cos (\Phi+\varphi)+\cos  \phi)\cos\varphi_\varepsilon}\ , \label{DF-spectral}
  \end{equation}
  with $\varphi_\varepsilon=\varepsilon/E_{Th}$ being the dynamical phase, accumulated by an excitation of the energy $\varepsilon$ on the closed path that  connects all four Y-junctions.   Note that the  Josephson term (\ref{jPhi}) does  not converge at high $\varepsilon$ and a regularization is needed. Indeed, the spectral current (\ref{DF-spectral}) depends periodically on the energy, due to the 1D nature 
  of the chiral modes carrying the current.   On physical grounds we expect this dependency to be replaced by 
  a slowly decaying (and oscillating) one, once the energy $\varepsilon$ reaches the lowest border of the 2D continuum, 
  ${\rm  min}(\Delta,M) \gg E_{Th}$. We, thus, smoothly cut off the integration in (\ref{jPhi}) at $\varepsilon \gg E_{Th}$ which leads to  the following current-phase relationship for equal   $T_l=T_r=T$
     	\begin{multline}
     	j_\Phi  =  4\pi  {\frac{e k_{\rm B} T}{h} }\sin(\Phi + \varphi) \times  \\ \sum \limits_{n=0}^\infty \frac{1}{2\exp\left(\pi \frac{k_{\rm B}T(1+2n)}{E_{Th}}\right)-\cos \phi-\cos (\Phi + \varphi) }. \label{CPhR}
     	\end{multline}
 The phase $\phi$ shifts $h/e$-periodic pattern of  critical current-flux oscillations, while $\varphi$ results into non-zero Josephson current without phase bias.
 The result for different  temperatures  $T_l\neq T_r$ equals half the sum of the two expressions (\ref{CPhR}) taken at $T=T_l$ and at $T=T_r$ respectively.

  To further support the validity of this the regularization procedure based on smooth cutoff, we can perform the derivation in a slightly different way that leads to the same result. Specifically, this alternative -- but equivalent -- regularization procedure amounts to subtracting and adding a high-temperature Josephson current at $T_l=T_r\gg E_{Th}$. 
  The difference of the Josephson current and the counter-term converges. At the same time, the counter-term is expected to be negligible on physical grounds. Indeed, the Josephson effect is possible due to the particle-hole coherence between the two Dirac channels. Such coherence extends to the length scales of the order of the thermal length $hv/(k_{\rm B} T)$. For $T\gg E_{Th}$ the thermal length is much shorter than the distance $L$ between the superconductors. Thus, the high-temperature Josephson effect is suppressed. This agrees with the results for S/N/S structures, where the normal region is a long quantum wire  \cite{Winkelholz,Maslov,Fazio,Dubos,Levchenko}. We emphasize once again that the convergence subtlety discussed here relates to the Josephson current only. The thermoelectric and the heat currents discussed below are 
  given by convergent integrals.

\section{Thermoelectric current}
 
 Below  we focus on the thermoelectric effect. Thus, we take  $T_l\neq T_r$ and set the SC phase bias  $\Phi=-\varphi$ so that the Josephson current $j_\Phi=0$ in (\ref{jFull}).
  Recall that we have redefined the Aharonov-Bohm phase $\phi$ in Eq. (\ref{phi}). We investigate the current $j_t$ as a function of two temperatures, $T_{l,r}$, and of $\phi$.
 We use the scattering matrices to express the  $\psi_{a,\varepsilon}$-operator  at $\Phi=-\varphi$ in the center of the upper Dirac channel  as 
   We use the expression (\ref{psi-a-0}) for $\psi_{a,\varepsilon}$-operator  at $\Phi=-\varphi$ in the center of the upper Dirac channel
   \begin{multline}
   \psi_{a,\varepsilon}=i \sqrt{2}  e^{\frac{1}{4} i \left(\frac{2 L \varepsilon}{v}+\phi \right)}\\ \frac{   \left(e^{i \varphi _{\varepsilon }}+e^{i \left(\varphi _{\varepsilon }+\phi \right)}-2 e^{i \phi }\right)\chi _l-\left(e^{i \phi }-1\right)e^{\frac{1}{2} i \left(\varphi _{\varepsilon }+\phi \right)} \chi _r  }{e^{i \varphi _{\varepsilon }} (1+e^{i \phi})^2 -4 e^{i \phi }}. \label{psi-a}
   \end{multline}
 The   $\psi_{b}$ is given by the interchanged  $\chi_{l}$ and $\chi_r$ for the rectangular geometry of the N-region.

\begin{figure}[h]
   	\includegraphics[width=\linewidth]{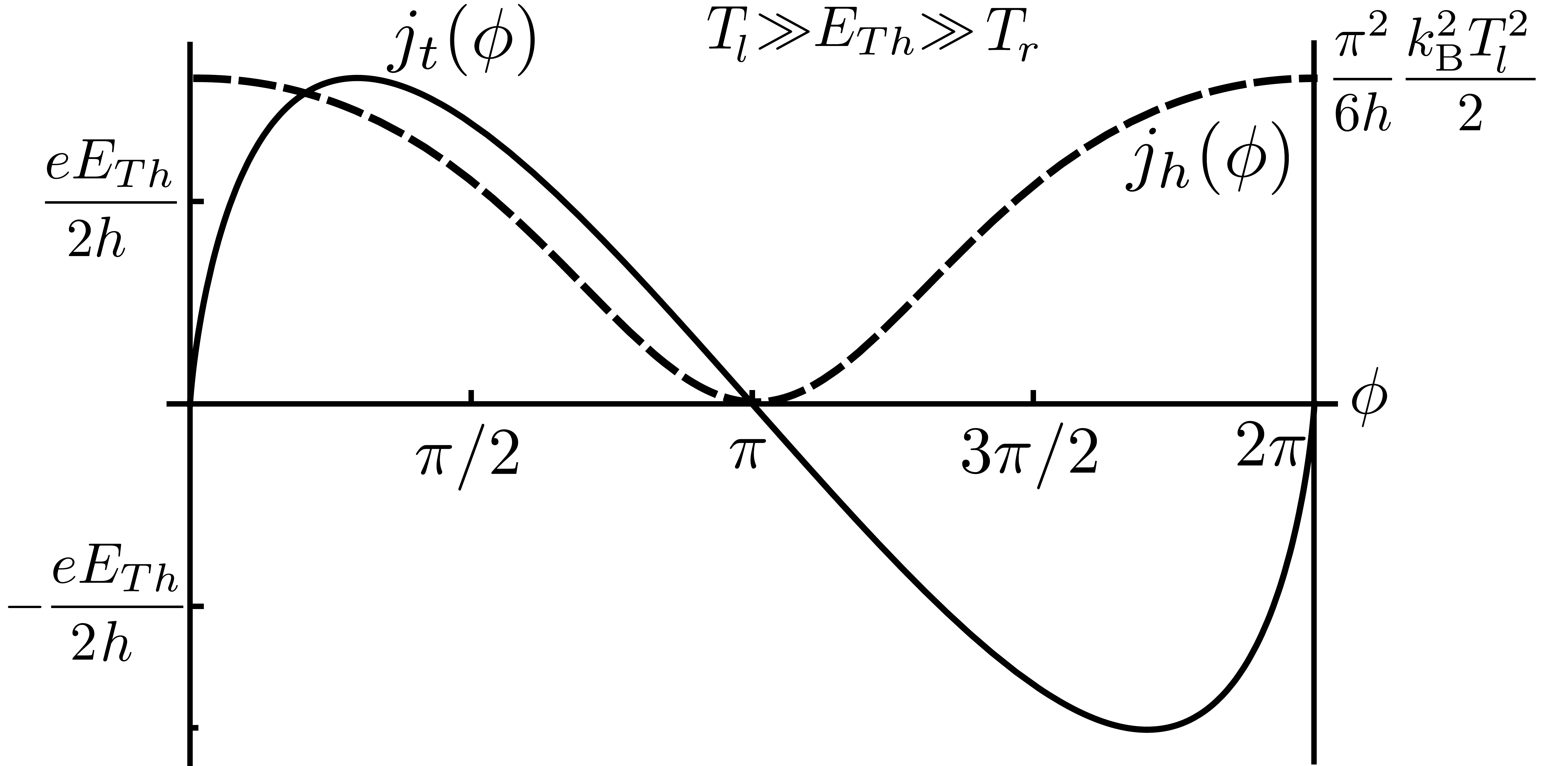}
    \caption{The thermoelectric current $j_t(\phi)$ and the heat current $j_h(\phi)$ as functions of the flux $\phi$ at $T_l\gg E_{Th}\gg T_r$.  } \label{j-phi}
   \end{figure}

 The above operator relations (\ref{psi-a}) allow the calculation of the   dimensionless spectral current $J_{t,\varepsilon}$ entering $j_t$, Eq. (\ref{j-t}):
      \begin{equation}
     J_{t,\varepsilon}
     = \frac{(1+\cos\phi)(1-\cos\varphi_\varepsilon) - \sin\phi\sin\varphi_\varepsilon}{1+\left(\frac{1+\cos\phi}{2}\right)^2-(1+\cos\phi)\cos\varphi_\varepsilon}. \label{j-E}
     \end{equation}
   The  term $(1+\cos\phi)(1-\cos\varphi_\varepsilon)$ in (\ref{j-E}) is an even function of $\varepsilon$ and does not contribute to the integral (\ref{j-t}) which is evaluated  by means of the summation over the residues of $\tanh\left( \varepsilon/(2T_{l,r})\right)$. Finally,  for arbitrary temperatures and $\phi$ we obtain an expression for the thermoelectric current
      \begin{multline}
   j_{  t} =\frac{ k_{ \rm B} e}{  \hbar}  \sin\phi \sum\limits_{n=0} \left(\frac{T_r}{2\exp\left(\frac{\pi  k_{ \rm B}T_r(1+2n)}{E_{Th}}\right)-1-\cos\phi}\right. \\
   \left.  -\frac{T_l}{2\exp\left(\frac{\pi  k_{ \rm B}T_l(1+2n)}{E_{Th}}\right)-1-\cos\phi}\right).\label{j-tl-tr}
   \end{multline}
   We focus on the regime of $T_l\gg E_{Th}\gg T_r$ or $T_r\gg E_{Th}\gg T_l$. We expect the maximal  current to be achieved in that region (see the solid curve in   Fig. \ref{j-phi}). In those cases  the  terms in the sum (\ref{j-tl-tr}) with the higher of the two temperatures are exponentially small. The sum of the remaining terms reduces to the integral  $T\sum\limits \to E_{Th}\int dx$, where $ x=T/E_{Th}$ and $T$ is the smaller of $T_r$ and $T_l$. The integral is proportional to $eE_{Th}/\hbar$:  
   \begin{equation}
   j_{t,max}={\sign} (T_l - T_r) \left. \frac{ e  E_{Th}}{2\pi \hbar}\tan\frac{\phi}{2} \, \ln\frac{2}{1-\cos\phi}\right. .
   \end{equation}
   Note the divergent derivative $\partial_\phi j_{max}(\phi)$ at $\phi=2\pi n$.
   In the high temperature regime, where $T_l,T_r\gg E_{Th}$, the thermoelectric current is exponentially suppressed and exhibits a sinusoidal dependence on $\phi$
   \begin{equation}
   	   	j =\frac{k_{\rm B} e}{2\hbar}  \sin\phi \left(T_r e^{-\pi T_r/E_{Th}}-T_l e^{-\pi T_l/E_{Th}}\right).
  \end{equation}   
    Similar to the  electric current, the thermoelectric current decays exponentially  at $k_{\rm B}T_{l,r}\gg E_{Th}$. This is the limit where the thermal length becomes much less than  the interferometer size.

    We next briefly address a general situation with nonzero Josephson and thermoelectric currents.
We derive from (\ref{psi-a-0}) that $J_{t,\varepsilon}$, entering the thermoelectric contribution (\ref{j-t}), reads for arbitrary temperatures, superconducting phases, and   Aharonov-Bohm phases as
  \begin{multline}
  	J_{t,\varepsilon}(\Phi)=\\ =
  	\frac{1+	(\cos\phi-\cos\varphi_\varepsilon)\cos (\Phi+\varphi)-\cos(\phi-\varphi_\varepsilon)  
  	}{1+\left(\frac{\cos  \phi +\cos (\Phi+\varphi)}{2}\right)^2-(\cos (\Phi+\varphi)+\cos  \phi)\cos\varphi_\varepsilon}\ , \label{J-E-phi}
  \end{multline}
Comparing the above equation (\ref{J-E-phi}) with $J_\varepsilon$ from (\ref{jPhi})   we relate the thermoelectric current $j_t$ and the Josephson currents $j_\Phi(T_{r,l})$ (\ref{CPhR}):
   \begin{equation}
   	j_t =\frac{\sin\phi }{2\sin(\Phi+\varphi)}\left( j_\Phi(T_r)-j_\Phi(T_l) \right). \label{j-t-1}
   \end{equation} 
At $T_r\ne T_l$, the Josephson current reads
   \begin{equation}
   	j_\Phi =\frac{1}{2 }\left( j_\Phi(T_r)+j_\Phi(T_l) \right).\label{j-phi-1}
   \end{equation} 
   
\begin{figure}[h]
   	\includegraphics[width=\linewidth]{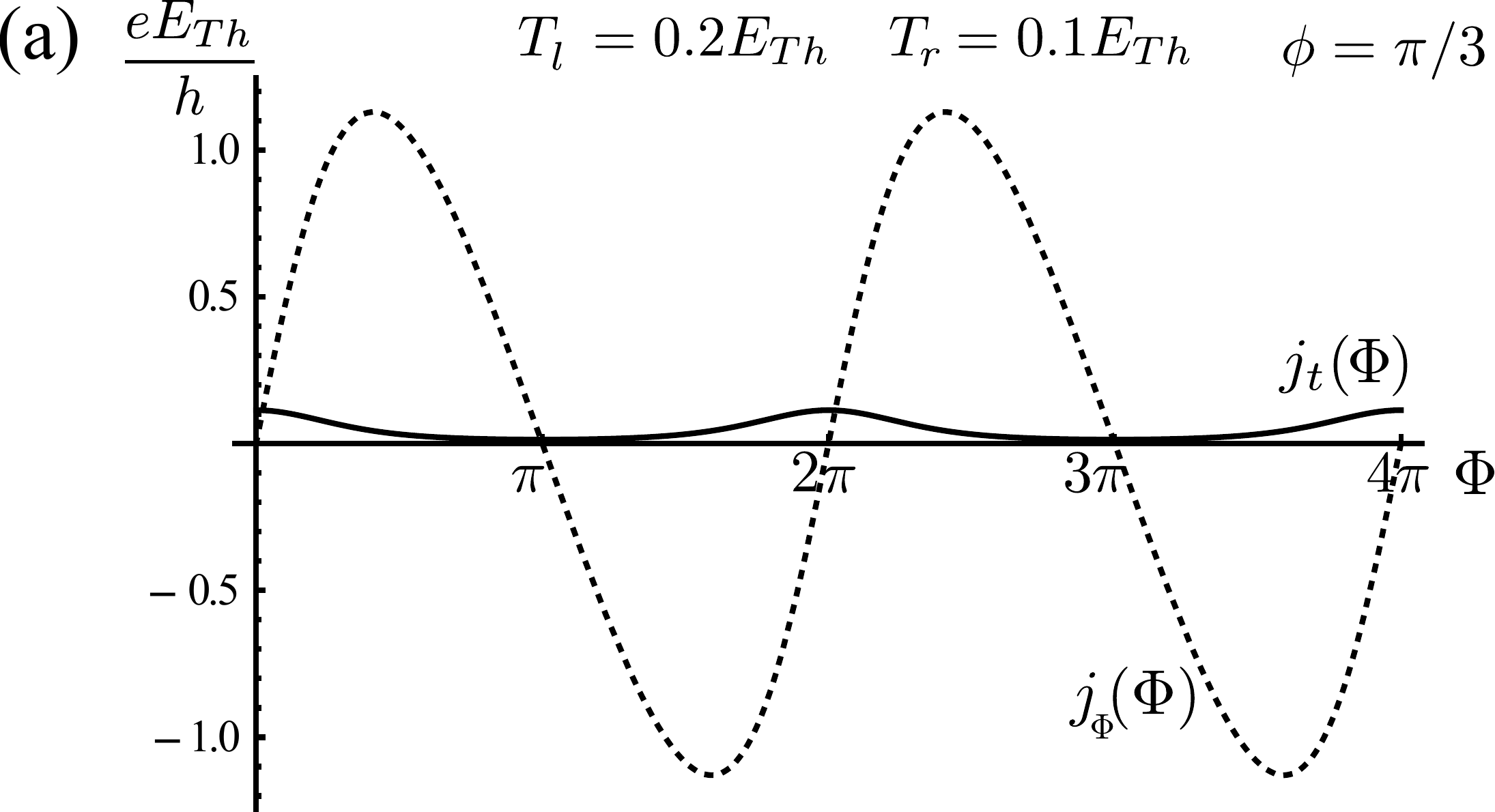}\\
   	\vspace{0.2cm}
   		\includegraphics[width=\linewidth]{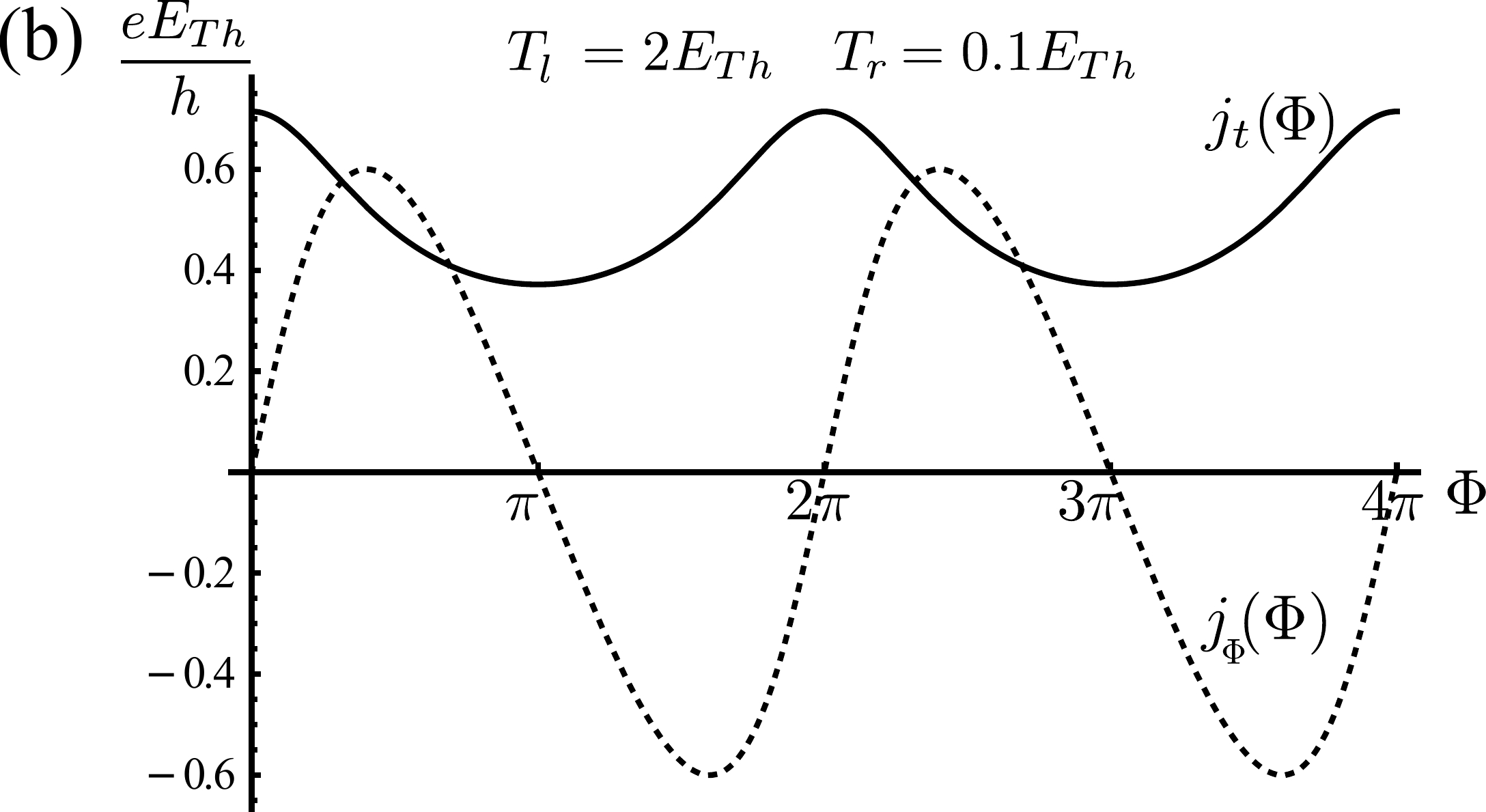}\\
   		\vspace{0.2cm}
   			\includegraphics[width=\linewidth]{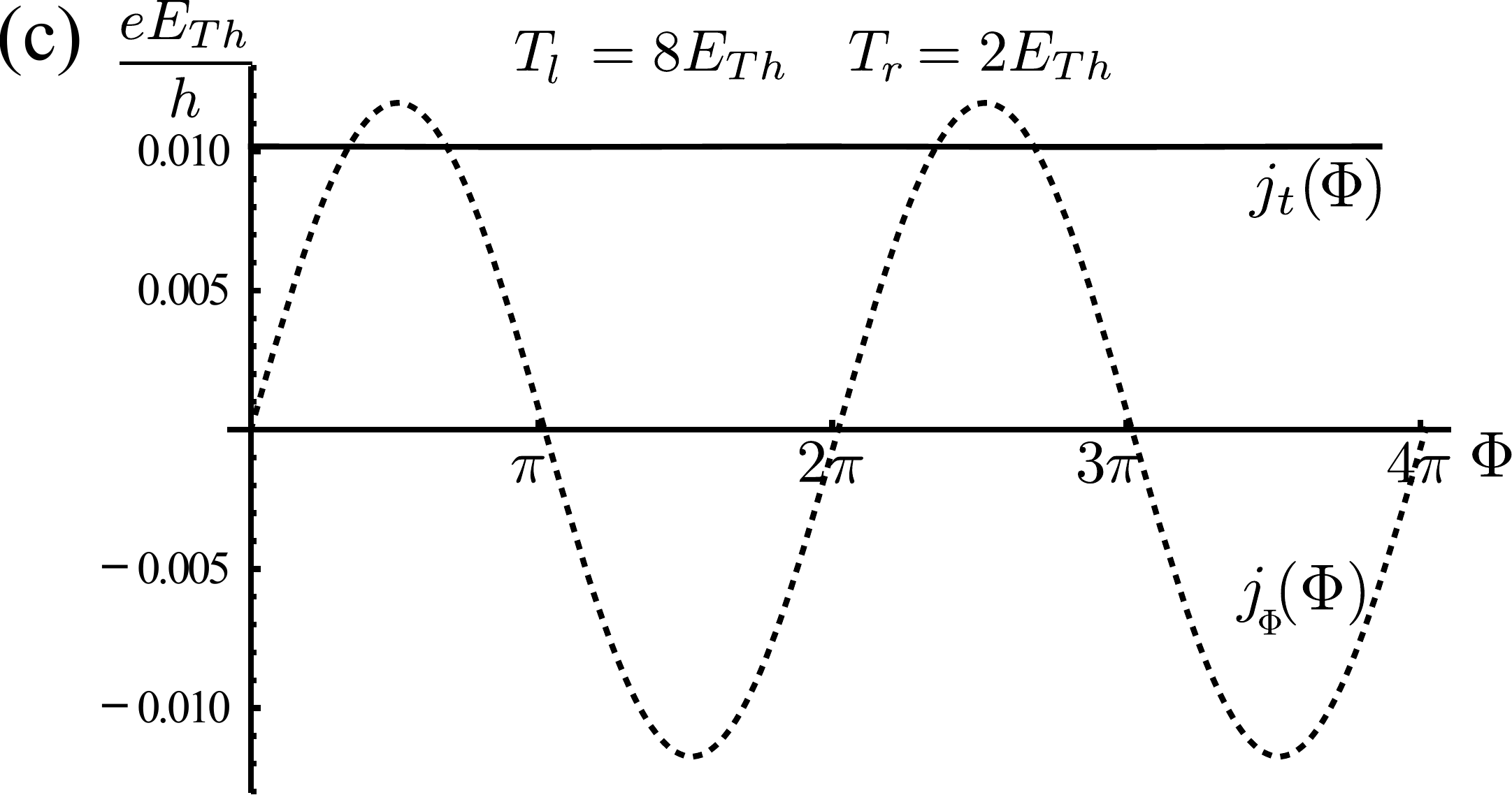}
   	\caption{The thermoelectric $j_t(\Phi)$ and Josephson $j_\Phi(\Phi)$ parts of the current as functions of the superconducting phase $\Phi$,   with the phase $\varphi$ set to zero. The Aharonov-Bohm phase is $\phi=\pi/3$.} \label{jt-jPhi-fig}
   \end{figure}

Fig. (\ref{jt-jPhi-fig}) shows the bias dependencies of the thermoelectric and Josephson currents at the Aharonov-Bohm phase $\phi=\pi/3$  in three temperature domains: (a) $T_{r,l}<E_{Th}$, (b)  $T_l>E_{Th}>T_r$   and (c) $T_{r,l}>E_{Th}$. In regime (a) the Josephson part is maximal, but the thermoelectric effect is suppressed. In (b,c) the thermal and Josephson parts are of the same orders of magnitude, $j_t\sim j_\Phi$. From (c) we see that the dependence of $j_t$ on $\Phi$ vanishes at   high temperatures.
   
Below we compute the bias phase $\Phi^*$ which results in zero total current 
  \begin{equation}
  	j(\Phi^*, T_r, T_l)=0.\label{j-2}
  	  \end{equation}
This value of the phase can be seen as an analogue of thermovoltage in the Josephson effect. In this regime (\ref{j-2}), where the Josephson and thermal currents compensate each other, one finds
  \begin{equation}
  	j_{\Phi^*}=-j_t. \label{j-t-j-phi}
  \end{equation}
     From (\ref{j-t-1},\ref{j-phi-1}) and (\ref{j-t-j-phi}) we obtain an equation on $\Phi^*$
  \begin{equation}
  	\sin(\Phi^*+\varphi) = \frac{j_{\Phi^*}(T_l)-j_{\Phi^*}(T_r)}{j_{\Phi^*}(T_l)+j_{\Phi^*}(T_r)} \sin\phi, \label{phi-1}
  \end{equation}     
 where $j_\Phi$    was introduced in (\ref{CPhR}). 
 
 The relation between $\Phi^*$ and the temperature gradient $\Delta T=T_r-T_l$ is nonlinear.
 Let us consider several limiting cases of (\ref{phi-1}) and their solutions. The first one is the high-temperature limit with $T_r, T_l \gg E_{Th}$. In this case the currents are
   \begin{equation}
 j_{\Phi }(T_{l,r})=2\pi \frac{e k_{\rm B} T_{l,r}}{h} \exp\left(-\frac{\pi T_{l,r}}{E_{Th}}\right) \sin(\Phi +\varphi), \label{j-3}
   \end{equation} 
   and the solution for $\Phi^*$ reads  
  \begin{equation}
  	\Phi^*= \arcsin\left( \sin\phi \tanh\frac{\pi (T_{r}-T_l)}{2E_{Th}} \right)-\varphi \label{phi-star-1}
  \end{equation}   
 In the limits of $T_{l}\gg E_{Th} \gg T_{r}$ or $T_{r}\gg E_{Th} \gg T_{l}$, one of the currents in (\ref{phi-1}) is temperature-independent and is given by $j_\Phi\propto E_{Th}$, while the other is exponentially suppressed as in (\ref{j-3}). The result is  
  \begin{equation}
  	\Phi^* = \phi \ {\sign} (T_{r}-T_l)-\varphi . \label{phi-star-2}
  \end{equation}
In the low-temperature limit $T_r, T_l \ll E_{Th}$,    for small gradients $\Delta T \ll T \ll E_{Th}$, we obtain  that
  \begin{equation}
  		\Phi^* =\frac{\Delta T}{3} \frac{\pi^2 k_{\rm B}^2 T}{E^2_{Th}} \frac{(1+\cos\phi)\sin\phi}{(1-\cos\phi)^2\ln\frac{2}{1-\cos\phi}}-\varphi \label{phi-star-3}
  \end{equation}
  with $\Delta T=T_r-T_l.$

   \section{  Heat current } 
\begin{figure}[h]
   	\includegraphics[width=\linewidth]{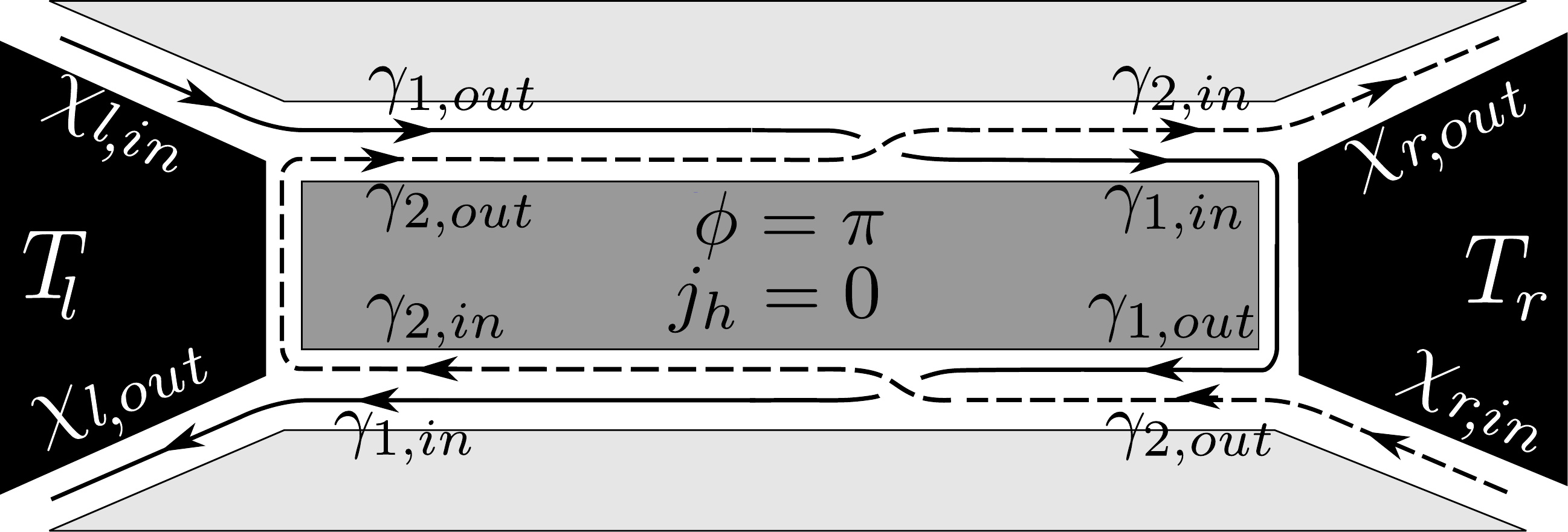}
 	\vspace{0.02\linewidth}\\
  	\includegraphics[width=\linewidth]{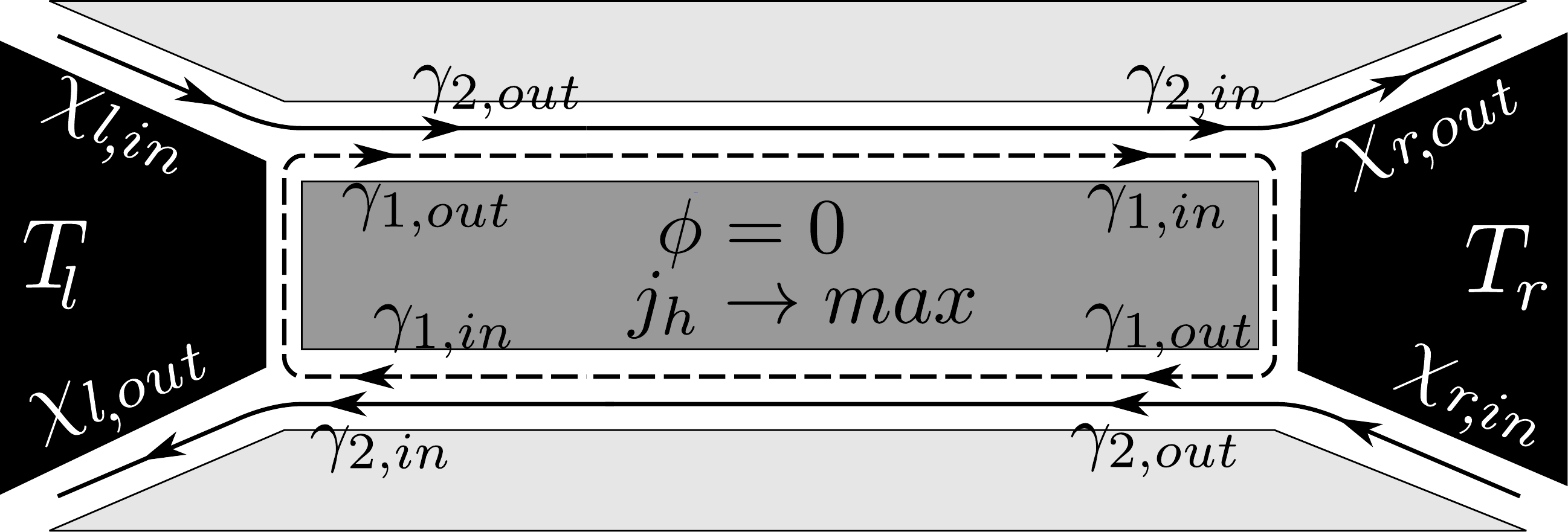}
   	\caption{Scattering in the basis of  the neutral modes $\chi, \gamma_1$ and $\gamma_2$ at a zero heat current (upper panel with $\phi=\pi$) and at the maximum of $j_h=\frac{\pi^2 k_{ \rm B}^2}{12 h } (T_l^2-T_r^2)$ (lower panel with $\phi=0$)} \label{setup-majorana-basis}
   \end{figure}

   The energy current at $\Phi=-\varphi$ is defined analogously to the thermoelectric one (\ref{j-t}) with the replacement of the electron charge  by the energy $(-e)\to \varepsilon$: 
   	\begin{equation}
    j_h= \int \frac{d\varepsilon}{2\pi \hbar}\varepsilon \frac{n_l(\varepsilon) - n_r(\varepsilon)}{2}J_{t,\varepsilon}. \label{jh0}
    \end{equation}
 The part in $J_{t,\varepsilon}$ (\ref{j-E}) which is proportional to $\sin\phi\sin\varphi_\varepsilon$ and contributes to $j_t$ does not contribute to $j_h$ while the term $\sim (1+\cos\phi)(1-\cos\varphi_\varepsilon)$ from $J_{t,\varepsilon}$ does contribute to the energy current. The result   of the integration  at an arbitrary temperatures  reads
   \begin{multline}
   j_h=\frac{\pi^2 k_{ \rm B}^2}{6 h }\frac{1+\cos\phi}{3+\cos\phi}(T_l^2-T_r^2) 
   +	\pi^2 \frac{1-\cos^2\phi }{3+\cos\phi} \times \\ \!\!\! \sum \limits_{n=0}^{\infty} \!\left( \frac{k_{ \rm B}^2T_r^2(2n+1)/h}{\exp\left[\frac{\pi k_{ \rm B} T_r(1+2n)}{E_{Th}}\right]-\cos^2(\phi/2)} - 
   (T_r{\to} T_l)
     \! \right). \label{j-h-1}
   \end{multline}	
The first term in (\ref{j-h-1}) gives a ballistic contribution to the heat conductance modulated by $\phi$. The heat current is  $h/e$-periodic like the electric current  but the energy current has always the same sing in contrast to the electric current $j_t$.  In the limit of $T_l\gg E_{Th}\gg T_r$ the amplitude of the heat current oscillations is maximal:  
  \begin{multline}
  	j_{h, max} = \frac{\pi^2 k_{ \rm B}^2T_l^2}{6 h}\frac{1+\cos\phi}{3+\cos\phi}  \\ + \frac{E_{Th}^2}{h}\frac{1-\cos\phi }{3+\cos\phi}  {\rm Li}_2 (\cos^2(\phi/2)),
  	\end{multline}
   where  ${\rm Li}_2 (z)$ is the polylogarithmic  function ${\rm Li}_n (z)=\sum\limits_{k=1}^\infty z^k/k^n.$ The dependence of $j_h(\phi)$ on the flux is shown in  Fig. \ref{j-phi} as a dashed curve.    Depending on Aharonov-Bohm phase the thermoelectric and heat currents can flow  in opposite ($0<\phi<\pi$) or in the same ($\pi<\phi<2\pi$) directions. One sees that the heat current  $j_h$ is maximal at $\phi=0$ with the value 
\begin{equation}
   j_{h}(0)=\frac{\pi^2 k_{ \rm B}^2T_l^2}{12 h }(T_l^2-T_r^2),
\end{equation}
which is half of the ballistic heat current of complex 1D fermions. The heat current is zero at $\phi=(2n+1)\pi$.
   The origin of the zeros of $j_h$ becomes transparent after one represents the Dirac $\psi$-operators in the normal region using a Majorana basis $\gamma_1$, $\gamma_2$: 
\begin{equation}
   	\gamma_1=(\psi+\psi^+)/\sqrt{2}, \quad \gamma_2=-i(\psi^+-\psi )/\sqrt{2}.
\end{equation}   	
The total phase $\phi=\phi_{AB}+\sum\limits_i^4\alpha_i=\pi$ can be interpreted as the sum of a zero Aharonov-Bohm phase and a set of redefined $\alpha_i$'s, for instance, with $\alpha_l=\alpha_1=\alpha_2=\pi/2$ and $\alpha_r=\alpha_3=\alpha_4=0$. The scattering between $\chi,\gamma_1$ and $\gamma_2$ for such contacts with equal phases $\alpha$ of Y-junctions was found in Ref. \onlinecite{ShapiroShnirmanMirlin}, Eq. (37)
   \begin{equation}
   \begin{bmatrix}
   \gamma_{1, out } \\ \\
   \chi_{_\beta,out}\\ \\
   \gamma_{2, out}
   \end{bmatrix}=
   \begin{bmatrix}
   \cos^2\alpha_\beta && -\sin\alpha_\beta && -\frac{\sin 2\alpha_\beta}{2} \\ \\
   -\sin\alpha_\beta && 0 && -\cos\alpha_\beta \\ \\
   \frac{\sin2\alpha_\beta}{2} && \cos\alpha && -\sin^2\alpha_\beta
   \end{bmatrix}
   \begin{bmatrix}
   \gamma_{1, in } \\ \\
   \chi_{_\beta,in}\\ \\
   \gamma_{2, in }
   \end{bmatrix} 
   \end{equation}
     with the index $\beta=l,r$. 
   From this representation of the scattering matrices for the left and right contacts, i.e. for $\alpha_l=\pi/2$ and $\alpha_r=0$, we find the paths of the scattered neutral modes  as shown in Fig. \ref{setup-majorana-basis}. 
   The incoming $\chi_{l,in}$ mode in the left lead converts into $\gamma_1$, propagates to the right lead, scatters in the normal region and flows back to the  left edge. There is no mixing between $\gamma_1$ (solid curve) and $\gamma_2$ (dashed curve) and no energy exchange between the two SCs. Hence, the heat current is zero. In the opposite case of the maximal heat current  which is equivalent to $\alpha=\alpha_l=\alpha_r=0$, the mode $\chi_{l,in}$ converts into  $\gamma_2$ in the left lead and flows away as $\chi_{r,out}$ in the right one.
   
   \section{Conclusions}
   We have analyzed the thermoelectric and heat transport  in a long 1D ballistic Josephson junction where the leads are formed by gapless chiral Majorana channels. Such a junction can be realized  as a hybrid structure based on a 3D topological insulator surface in the proximity with s-wave superconducting and magnetic films. The interfaces of the gapped sectors support  neutral and charged 1D  chiral liquids. The normal region  is formed by two chiral Dirac liquids spaced by a magnetic material. The chiral contact is formed by four Y-junctions which serve as Dirac-Majorana converters. Our crucial assumption is that the Thouless energy, proportional  to the inverse dwell time in the interferometer, is much lower than the superconducting and exchange gaps.
   
 We have obtained the following results. (i) We have generalized the current-phase relation from our previous work ~\cite{ShapiroShnirmanMirlin} to nonidentical Dirac-Majorana contacts and  discovered a nonzero Josephson current in the absence of the phase bias, the Aharonov-Bohm phase, and temperature gradient. (ii) We have calculated the thermoelectric current and (iii) the heat current as functions of the magnetic flux and the temperatures of the leads. 
 An important difference of the chiral contact from junctions based on a 2D TI, quantum Hall bar or a spin-orbit coupled nanowire  is the absence of a quasiparticle gap in the leads due to the gapless Majorana modes.  This  results in  the absence of the temperature threshold  in the current-flux relations.  We observe a  non-sinusoidal $2\Phi_0 $-periodic dependence of the thermoelectric and heat currents on the magnetic flux. The maximum oscillation amplitude of the thermoelectric current is proportional to $eE_{Th}/\hbar$  and scales as one over the device size. The maximal amplitude is achieved at a low temperature in one of the superconductors and a high temperature in the other one,  i.e. at $T_{l}\gg E_{Th} \gg T_{r}$ or $T_{r}\gg E_{Th} \gg T_{l}$.    The heat current oscillates between zero and a value that corresponds to one half of the heat conductance quantum.

\section{   Acknowledgments } This research was financially supported by the DFG-RSF grant (No. 16-42-01035 (Russian node) and No. SH 81/4-1, MI 658/9-1  (German node)). DEF's research was supported in part by the National Science Foundation under Grants No. DMR-1607451 and PHY-1125915.

\end{document}